\documentclass[a4paper,12pt]{article}
\usepackage{jcappub}
\usepackage[symbol]{footmisc}
\usepackage{amsmath,amssymb}
\usepackage{revsymb}
\usepackage{orcidlink}
\newcommand{\nc}{\newcommand*}

\nc{\Eq}[1]{Eq.~\eqref{#1}}     % equation
\nc{\Fig}[1]{Fig.~\ref{#1}}     % figure
\nc{\Table}[1]{Table~\ref{#1}}  % table
\nc{\Sec}[1]{Sec.~\ref{#1}}     % section
\def\({\left(}
\def\){\right)}
\def\[{\left[}
\def\]{\right]}
\def\e{\begin{equation}}
\def\q{\end{equation}}
\def\m{\begin{eqnarray}}
\def\n{\end{eqnarray}}

\begin{document}

\title{Constraining string cosmology with the gravitational-wave background using the NANOGrav 15-year data set}

%%%%%%%%%%%%%%%%%%%%%%%%%%%%%%%%%%%%%%%%%%%%%%%%%%%%%%%%%%%%%%%%%%%%%%%%%%%%%%%%%%%%%%%%%%%%%%%%
\author[a,b]{Qin Tan,\orcidlink{0000-0002-9496-6476}}
\author[c,*]{You~Wu,\note{Corresponding author.}\orcidlink{0000-0002-9610-2284}}
\author[d,*]{Lang~Liu\orcidlink{0000-0002-0297-9633}}

\affiliation[a]{Department of Physics, Key Laboratory of Low Dimensional Quantum Structures and Quantum Control of Ministry of Education, Synergetic Innovation Center for Quantum Effects and Applications, Hunan Normal University, Changsha, 410081, Hunan, China}
\affiliation[b]{Institute of Interdisciplinary Studies, Hunan Normal University, Changsha, Hunan 410081, China}
\affiliation[c]{College of Mathematics and Physics, Hunan University of Arts and Science, Changde, 415000, China}
\affiliation[d]{Faculty of Arts and Sciences, Beijing Normal University, Zhuhai 519087, China}

\emailAdd{tanqin@hunnu.edu.cn}
\emailAdd{youwuphy@gmail.com}
\emailAdd{liulang@bnu.edu.cn}

%%%%%%%%%%%%%%%%%%%%%%%%%%%%%%%%%%%%%%%%%%%%%%%%
\abstract{The North American Nanohertz Observatory for Gravitational Waves (NANOGrav) collaboration has recently reported strong evidence for a signal at nanohertz, potentially the first detection of the stochastic gravitational-wave background (SGWB). We investigate whether the NANOGrav signal is consistent with the SGWB predicted by string cosmology models. By performing Bayesian parameter estimation on the NANOGrav 15-year data set, we constrain the key parameters of a string cosmology model: the frequency $f_s$ and the fractional energy density $\Omega_\mathrm{gw}^{s}$ of gravitational waves at the end of the dilaton-driven stage, and the Hubble parameter $H_r$ at the end of the string phase. Our analysis yields constraints of $f_s = 1.2^{+0.6}_{-0.6}\times 10^{-8} \mathrm{Hz}$ and $\Omega_\mathrm{gw}^{s} = 2.9^{+5.4}_{-2.3}\times 10^{-8}$, consistent with theoretical predictions from string cosmology. However, the current NANOGrav data is not sensitive to the $H_r$ parameter. We also compare the string cosmology model to a simple power-law model using Bayesian model selection, finding a Bayes factor of $2.2$ in favor of the string cosmology model. Our results demonstrate the potential of pulsar timing arrays to constrain cosmological models and study the early Universe.}
	
\maketitle

%%%%%%%%%%%%%%%%%%%%%%%%%%%%%%%%%%%%%%%%%%%%%%%%%%%%%%%%%%%%%%%%%%%%%%%%%%%%%%%%%%%%%%%%%%%%%%%%
\section{Introduction}

Building upon the groundbreaking detection of gravitational waves (GWs) from the coalescence of black holes and neutron stars by LIGO-Virgo-KAGRA (LVK)~\cite{Abbott:2016blz,TheLIGOScientific:2017qsa,LIGOScientific:2018mvr,LIGOScientific:2020ibl,LIGOScientific:2021usb,LIGOScientific:2021djp}, the next exciting discovery may be the identification of the stochastic GW background (SGWB), which can span a wide frequency range. Pulsar timing arrays (PTAs) serve as indispensable tools for probing the nanohertz frequency band of the SGWB, offering a valuable window into the detection of GWs that originated from the early Universe. 

Recently, multiple PTA collaborations, including the North American Nanohertz Observatory for GWs (NANOGrav)~\cite{NANOGrav:2023gor,NANOGrav:2023hde}, the Parkes PTA (PPTA)~\cite{Zic:2023gta,Reardon:2023gzh}, the European PTA (EPTA) in partnership with the Indian PTA (InPTA)~\cite{EPTA:2023sfo,Antoniadis:2023ott}, and the Chinese PTA (CPTA)~\cite{Xu:2023wog}, have independently provided strong evidence for spatial correlations that are consistent with the Hellings-Downs~\cite{Hellings:1983fr} pattern in their most recent data sets. These correlations align with the expected properties of an SGWB as predicted by the theory of general relativity. These discoveries mark a pivotal achievement in the field of GW astronomy, as they demonstrate the successful detection of GWs through the meticulous timing of exceptionally stable millisecond pulsars.

Despite these remarkable achievements, the precise origin of the observed PTA signals remains uncertain~\cite{NANOGrav:2023hvm,Antoniadis:2023xlr}, with hypotheses encompassing both astrophysical and cosmological sources. The diverse range of potential origins includes supermassive black hole binaries (SMBHBs)~\cite{NANOGrav:2023hfp,Ellis:2023dgf,Shen:2023pan,Bi:2023tib,Barausse:2023yrx}, as well as more exotic phenomena such as scalar-induced GWs~\cite{Inomata:2023zup,Chen:2019xse,Liu:2023ymk,Franciolini:2023pbf,HosseiniMansoori:2023mqh,Wang:2023ost,Jin:2023wri,Liu:2023pau,Zhao:2023joc,Yi:2023npi,Harigaya:2023pmw,Balaji:2023ehk,Yi:2023tdk,You:2023rmn,Liu:2023hpw,Choudhury:2023fwk,Choudhury:2023fjs,Domenech:2024rks,Chen:2024twp,Choudhury:2024dzw,Choudhury:2024aji,Chen:2024fir}, cosmic phase transitions~\cite{Addazi:2023jvg,Athron:2023mer,Zu:2023olm,Jiang:2023qbm,Xiao:2023dbb,Abe:2023yrw,Gouttenoire:2023bqy,An:2023jxf,Chen:2023bms}, domain walls~\cite{Kitajima:2023cek,Blasi:2023sej,Babichev:2023pbf,Guo:2023hyp}, cosmic strings~\cite{Chen:2022azo,Kitajima:2023vre,Ellis:2023tsl,Wang:2023len,Ahmed:2023pjl,Antusch:2023zjk}, and modified gravities~\cite{Chen:2021wdo,Wu:2021kmd,Chen:2021ncc,Wu:2023pbt,Chen:2023uiz,Bi:2023ewq,Wu:2023rib}. Another intriguing possibility is that the detected signal may be the SGWB originating from string cosmology.

The standard cosmological model \cite{Coley:2019yov} has achieved great success in describing the behavior of our Universe. When augmented with the inflationary epoch \cite{Guth:1980zm}, this model offers a compelling explanation for conundrums such as the fine-tuning of initial conditions and demonstrates excellent concordance with the observed inhomogeneous structure of the Universe.  However, this mechanism is not without its limitations. In most models of inflation based on a scalar field minimally coupled to gravity, the inflationary period lasts so long that the physical fluctuations corresponding to the present large-scale structure would have shrunk to scales smaller than the Planck length at the beginning of inflation. This is known as the ``trans-Planckian"  problem \cite{Martin:2000xs}. Furthermore, as we look back in time, the spacetime curvature increases, inevitably leading to an initial singularity \cite{Borde:1993xh,Borde:2001nh} from the Big Bang. Additionally, our understanding of the physical nature of the inflation field remains limited due to its exotic characteristics.

Quantum effects of gravity are inescapable in the primordial Universe. Consequently, string theory may offer solutions to these challenges. The resulting string cosmology gives rise to a pre-big bang scenario \cite{Gasperini:1992em,Gasperini:2007vw} in which extra dimensions are introduced, and the small characteristic size of strings \cite{Kaplunovsky:1985yy} circumvents the initial singularity encountered in general relativity. In this framework, the Universe can commence inflation with a large Hubble horizon, thereby resolving the trans-Planckian problem. String theory predicts the presence of a scalar dilaton field coupled to gravity, which induces an inflationary process distinct from the standard slow-roll inflation \cite{Linde:1990ta}. As a fascinating consequence, pre-big bang inflation generates a primordial SGWB with a blue tilt \cite{Grishchuk:1991kk,Brustein:1995ah,Jiang:2023qht}. Thus, the SGWB arising from string cosmology provides a natural explanation for the recently observed PTA signal.

In this work, we assume that the signal detected by PTAs has its origin in string cosmology and employ the PTA observations to constrain the string cosmology model. Our primary objectives are to investigate whether the PTA signal is consistent with the string cosmology model and to place constraints on the parameter space of the string cosmology model. The remainder of the paper is structured as follows. Section \ref{SGWB} provides an overview of the SGWB in the context of the string cosmology scenario. In Section \ref{data}, we outline the methodology for data analysis and present the results obtained from the NANOGrav 15-year data set. Finally, we draw conclusions in Section \ref{conclusion}.

%%%%%%%%%%%%%%%%%%%%%%%%%%%%%%%%%%%%%%%%%%%%%%%%%%%%%%%%%%%%%%%%%%%%%%%%%%%%%%%%%%%%%%%%%%%%%%%%
\section{\label{SGWB}SGWB from string cosmology}
In this section, we provide a brief overview of the SGWB in the context of the string cosmology scenario. In a typical string cosmology model, the Universe undergoes two early inflationary phases: the ``dilaton-driven" stage and the ``string" stage~\cite{Brustein:1995ah}. Each stage generates a SGWB, which can be characterized using the spectral function $\Omega_{\text{GW}}(f)$, defined as
\begin{equation}
	\Omega_{\text{GW}}(f)=\frac{1}{\rho_{\text{c}}}\frac{d\rho_{\text{GW}}}{d\ln f},\label{Omega_function1}
\end{equation}
where $\rho_{\text{GW}}$ denotes the energy density of the SGWB between frequencies $f$ and $f+df$, and $\rho_{\text{c}}$ is the critical energy density, given by
	\begin{equation}
		\rho_{\text{c}}=\frac{3c^{2}H_{0}^{2}}{8\pi G}.\label{critical_energy_density}
		\end{equation}
According to~\cite{	Brustein:1996mq,Allen:1996sw}, the spectrum of the SGWB in the string cosmology scenario can be approximated as
		\begin{equation}
			\Omega_\text{GW}(f) = \begin{cases}
				\Omega_\text{gw}^s (f/f_s)^3, & f<f_s, \\
				\Omega_\text{gw}^s (f/f_s)^\beta, & f_s<f<f_1, \\
				0, & f_1<f,
			\end{cases}
			\label{Omega_spectrum}
		\end{equation}
where
\begin{equation}
	\beta = \frac{\log \left(\Omega_\text{gw}^\text{max}/\Omega_\text{gw}^s \right)}{\log \left( f_1/f_s\right)}
\end{equation}
is the logarithmic slope of the spectrum of the SGWB produced during the string epoch. The spectrum depends on four parameters: the frequency $f_{s}$, the fractional energy density $\Omega_{\text{gw}}^{s}$ generated at the end of the dilaton-drive stage, the maximum frequency $f_{1}$ (above which no gravitational radiation is generated), and the maximum fractional energy density $\Omega_{\text{gw}}^{\text{max}}$ occurring at the maximum frequency $f_{1}$. Assuming no late entropy generation and a reasonable choice about the number of effective degrees of freedom, we can express $f_{s}$ and $\Omega_{\text{gw}}^{\text{max}}$ in terms of the Hubble parameter $H_{r}$ at the end of the string phase as~\cite{Brustein:1996ut}
\begin{equation}
	f_1 = 1.3\times10^{10} \left( \frac{H_r}{5\times10^{17}\,\text{GeV}}\right)^{1/2}\,\text{Hz}, \label{eq:f_1}
\end{equation}
and 
\begin{equation}
	\Omega_\text{gw}^\text{max} = 1\times10^{-7} h_{100}^{-2}\left( \frac{H_r}{5\times10^{17}\,\text{GeV}}\right)^{2}, \label{eq:omega_max}
\end{equation}
where $h_{100}=0.674$~\cite{Planck:2018vyg} is the reduced Hubble constant. The spectrum is now determined by only three parameters, $f_{s}$, $\Omega_\text{gw}^s$, and $H_{r}$, which are related to the fundamental parameters of the string cosmology model. Although some studies~\cite{Buonanno:1996xc,Galluccio:1996xa} consider more complex models and their resulting spectra, the results are similar to the spectrum discussed here and equally dependent on the same parameters. Therefore, the model we use captures the main features of string cosmology. In the next section, we will use data from the NANOGrav to estimate the parameters of this string cosmology model. By employing Bayesian parameter estimation techniques, we aim to provide insights into the search for observational signatures of string cosmology.

%%%%%%%%%%%%%%%%%%%%%%%%%%%%%%%%%%%%%%%%%%%%%%%%%%%%%%%%%%%%%%%%%%%%%%%%%%%%%%%%%%%%%%%%%%%%%%%%
\section{\label{data}Data analysis and results}

In this work, we utilize the NANOGrav 15-year data set~\cite{NANOGrav:2023hde} to estimate the parameters of the string cosmology model. The NANOGrav 15-year data set includes observations of 67 millisecond pulsars with a timing baseline $\geq 3$ years, spanning a total time span of approximately $16.03$ years~\cite{NANOGrav:2023hde,NANOGrav:2023gor}. This extensive data set provides a unique opportunity to search for the SGWB signal and constrain cosmological models, such as string cosmology.

We specifically employ the free spectrum amplitudes obtained by the NANOGrav 15-year data set when considering spatial correlations of the Hellings-Downs pattern~\cite{Hellings:1983fr}. The Hellings-Downs correlation pattern is a distinctive signature of the SGWB, arising from the quadrupolar nature of GWs. By incorporating this pattern into our analysis, we can effectively distinguish the SGWB signal from other noise sources, such as intrinsic pulsar noise or Earth-term errors~\cite{Taylor:2021yjx}.

\begin{figure}[tbp!]
	\centering
 \includegraphics[width=\textwidth]{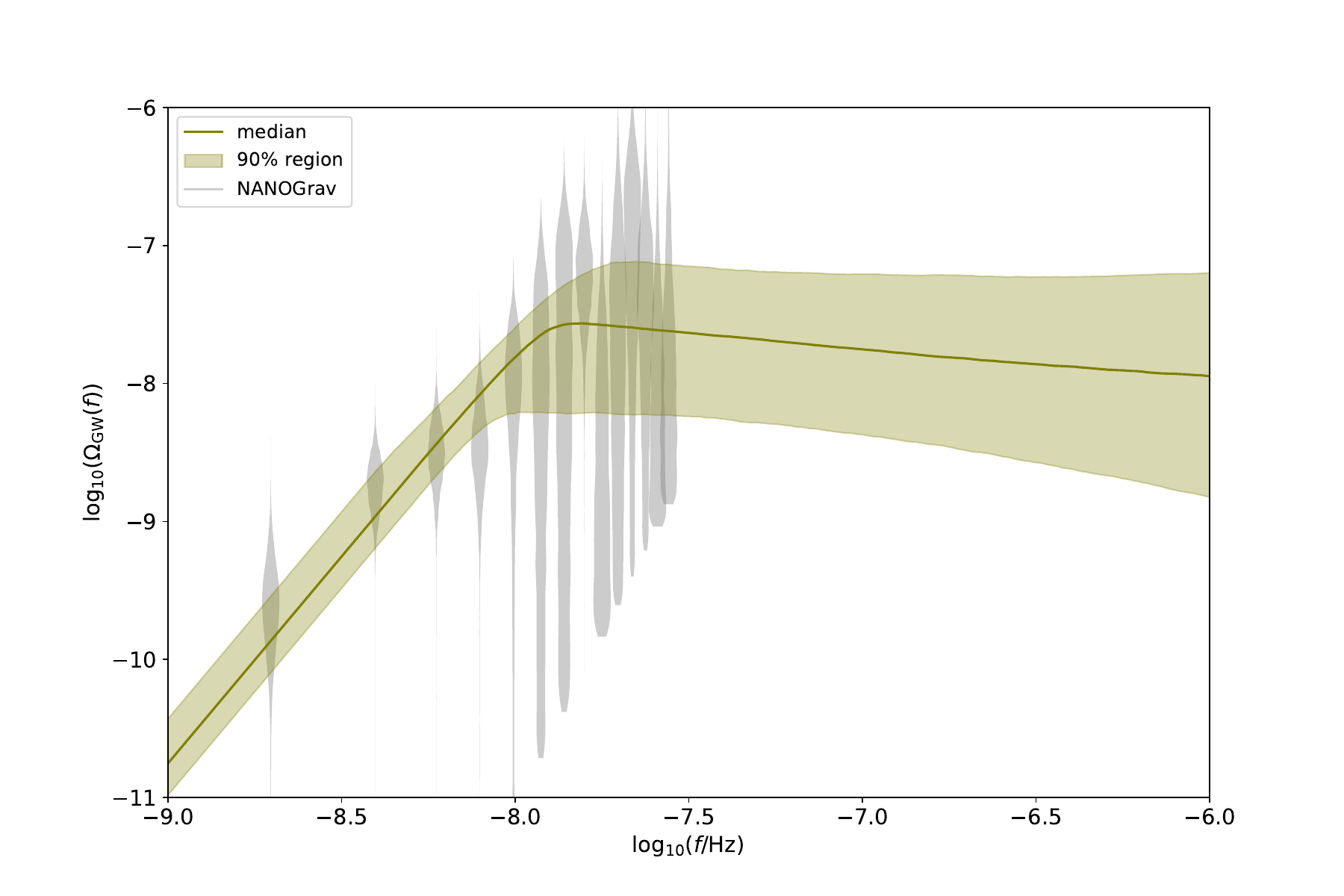}
	\caption{\label{ogw}PPD for the energy density spectrum of the SGWB from string cosmology. The gray violins represent the free spectra obtained from the NANOGrav 15-year data set, while the olive shaded region indicates the $90\%$ credible interval of the posterior distribution. The PPD showcases the compatibility of the string cosmology model with the observed data, highlighting the constraining power of the NANOGrav measurements on the model parameters}
\end{figure}

The sensitivity of a PTA's observations to the SGWB begins at a frequency of $1/T_{\mathrm{obs}}$, where $T_{\mathrm{obs}}=16.03\,\mathrm{yr}$ is the observational time span. NANOGrav employs $14$ frequency bins~\cite{NANOGrav:2023gor} in their search for the SGWB signal, covering a frequency range from $2.0 \times 10^{-9}$ Hz to $2.8 \times 10^{-8}$ Hz. This frequency range is particularly relevant for detecting the SGWB signal predicted by string cosmology models, as the signal is expected to have a significant contribution at frequencies around $10^{-8}$ Hz. In \Fig{ogw}, we illustrate the data employed in our analyses and depict the energy density.

In our analysis, we start by utilizing the posterior data of the time delay $d(f)$ obtained from the NANOGrav 15-year data set. The time delay is related to the power spectrum $S(f)$ through the following relation:
\begin{equation}
S(f) = d(f)^2 T_{\mathrm{obs}},
\end{equation}
where $T_{\mathrm{obs}}$ represents the total observational time span. This relation allows us to calculate the power spectrum from the time delay data, which is a crucial step in determining the energy density of the SGWB.

%%%%%%%%%%%%%%%%%%%%%%%%%%%%%%%%%%%%%%%%%%%%%%%%%%%%%%%%%%%%%%%%%
\begin{table}
    \centering
	\begin{tabular}{cccc}
		\hline
		Parameter & Prior & Result \\ 
  \hline \\[-2\medskipamount]   
  $f_s/\mathrm{Hz}$ & LogUniform$[10^{-10}, 10^{-5}]$ & $1.2^{+0.6}_{-0.6}\times 10^{-8}$\\[1pt]
  $\Omega_\mathrm{gw}^{s}$ & LogUniform$[10^{-8}, 10^{-7}]$ & $2.9^{+5.4}_{-2.3}\times 10^{-8}$\\[1pt]
  $H_r/\mathrm{GeV}$ & LogUniform$[10^{12}, 10^{19}]$ & -\\[1pt]
  \hline
	\end{tabular}
	\caption{\label{tab:priors}Prior distributions and posterior estimates for the string cosmology model parameters. The priors are specified as log-uniform distributions over the indicated ranges. The posterior estimates are reported as median values along with $90\%$ equal-tail credible intervals. The parameter $H_r$, representing the Hubble parameter at the end of the string phase, remains unconstrained by the NANOGrav data.}
\end{table}

%%%%%%%%%%%%%%%%%%%%%%%%%%%%%%%%%%%%%%%%%%%%%%%%%%%%%%%%%%%%%%%%%
\begin{figure}[tbp!]
	\centering
 \includegraphics[width=0.8\textwidth]{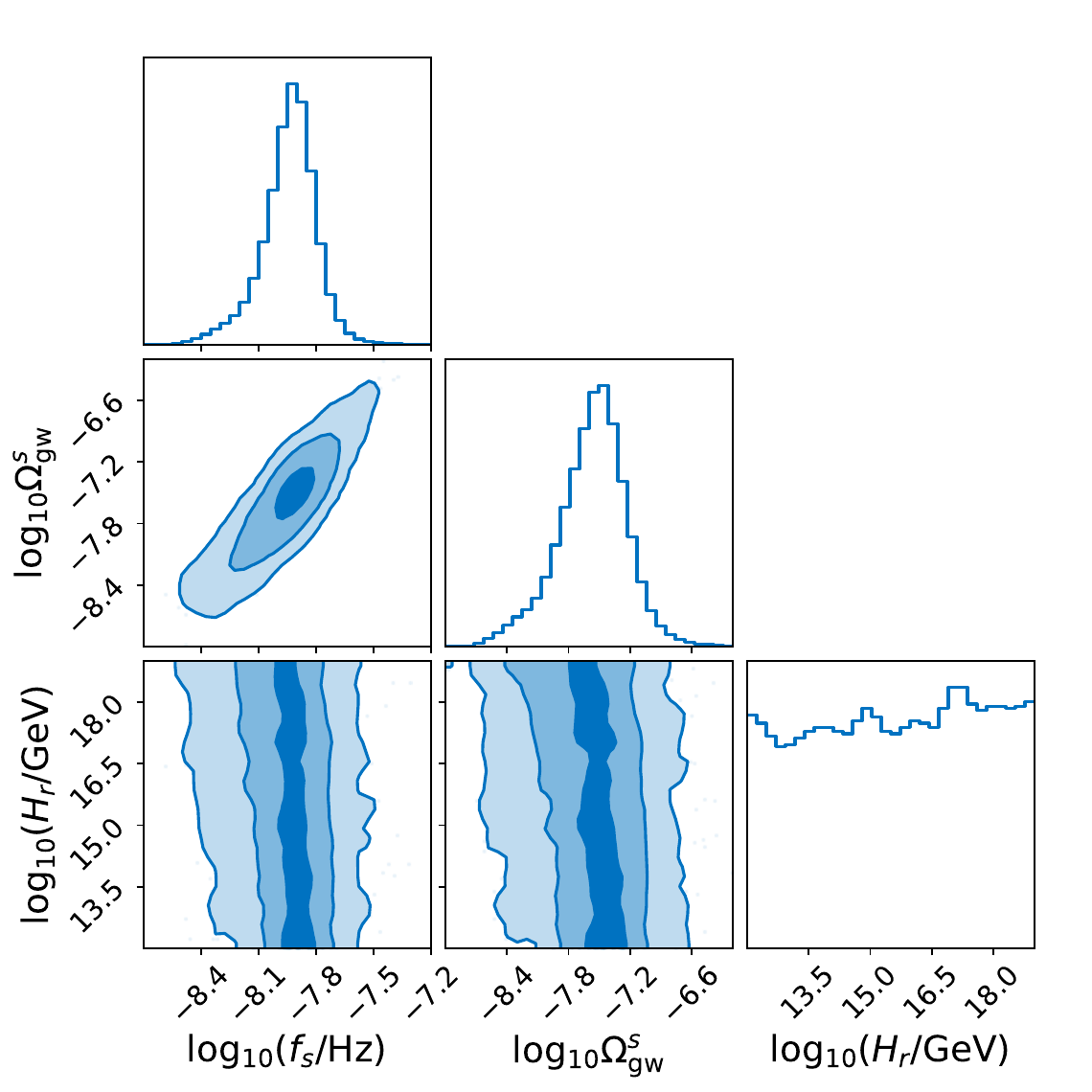}
	\caption{\label{posts} Marginal posterior distributions for the string cosmology model parameters, $\Lambda = \{f_s, \Omega_\mathrm{gw}^{s}, H_r\}$, derived from the NANOGrav 15-year data set. The one-dimensional histograms show the marginalized posteriors for each parameter, while the two-dimensional plots display the joint posterior distributions, with contours delineating the $1 \sigma$, $2 \sigma$, and $3 \sigma$ credible regions.}
\end{figure}

Using the power spectrum, we can then compute the energy density of the free spectrum, $\hat{\Omega}_{\mathrm{GW}}(f)$, as
\begin{equation}
\hat{\Omega}_{\mathrm{GW}}(f) = \frac{2 \pi^2}{3 H_0^2} f^2 h_c^2(f) = \frac{8\pi^4}{H_0^2} T_{\mathrm{obs}} f^5 d^2(f),
\end{equation}
where $H_0 = 67.4\, \mathrm{km}\,\mathrm{s}^{-1}\,\mathrm{Mpc}^{-1}$ is the Hubble constant determined by the Planck collaboration~\cite{Planck:2018vyg}. Here, the characteristic strain, $h_c(f)$, is defined as
\begin{equation}
h_c^2(f) = 12 \pi^2 f^3 S(f).
\end{equation}
By combining the power spectrum and the characteristic strain, we can fully characterize the SGWB and its energy density.

For each of the 14 observed frequencies $f_i$ in the NANOGrav data set, we estimate the corresponding kernel density, $\mathcal{L}_i$, using the obtained posteriors of $\hat{\Omega}_{\mathrm{GW}}(f_i)$. The total log-likelihood is then calculated as the sum of the individual log-likelihoods for each frequency, as~\cite{Moore:2021ibq,Lamb:2023jls,Liu:2023ymk,Wu:2023hsa,Jin:2023wri,Liu:2023pau}
\begin{equation}
\ln \mathcal{L}(\Lambda) = \sum_{i=1}^{14} \ln \mathcal{L}_i(\Omega_{\mathrm{GW}}(f_i, \Lambda)),
\end{equation}
where $\Lambda = \{f_s, \Omega_\mathrm{gw}^{s}, H_r\}$ represents the set of three model parameters that we aim to constrain using the NANOGrav data. These parameters are the frequency $f_s$ and the fractional energy density $\Omega_\mathrm{gw}^{s}$ of GWs generated at the end of the dilaton-driven stage, and the Hubble parameter $H_r$ at the end of the string phase.

To efficiently explore the parameter space and obtain posterior distributions for the model parameters, we employ the \texttt{dynesty}~\cite{Speagle:2019ivv} sampler, which is a nested sampling algorithm implemented in the \texttt{Bilby}~\cite{Ashton:2018jfp,Romero-Shaw:2020owr} package. Nested sampling is a powerful technique for Bayesian inference, as it allows for the computation of the evidence (marginal likelihood) and the posterior distributions simultaneously. The priors and results for the model parameters are summarized in~\Table{tab:priors}, providing an overview of the constraints obtained from our analysis of the NANOGrav 15-year data set.

%%%%%%%%%%%%%%%%%%%%%%%%%%%%%%%%%%%%%%%%%%%%%%%%%%%%%%%%%%%%%%%%%

We present the posterior distributions for the model parameters in \Fig{posts}. Our analysis reveals that to explain the PTA signal detected by the NANOGrav 15-year data set, the parameters should satisfy $f_s = 1.2^{+0.6}_{-0.6}\times 10^{-8} \mathrm{Hz}$, $\Omega_\mathrm{gw}^{s} = 2.9^{+5.4}_{-2.3}\times 10^{-8}$, and $H_r$ shows no constraints, following the prior distribution. This indicates that the current data is not sensitive to the $H_r$ parameter.
The posterior predictive distribution (PPD) of our model is shown in \Fig{ogw}, demonstrating the agreement between our model and the observed NANOGrav data. Specifically, the PPD is consistent with the free spectrum amplitudes obtained by NANOGrav, indicating that our string cosmology model provides a good fit to the data.

To assess the relative performance of our string cosmology model compared to alternative explanations for the NANOGrav signal, we have also calculated the Bayes factor between the string cosmology model and a simple power-law model, which is commonly associated with the SGWB from SMBHBs~\cite{Sesana:2008mz,Chen:2018rzo}. The Bayes factor is defined as the ratio of the marginal likelihoods of two competing models, and it quantifies the relative support for each model given the observed data~\cite{Kass:1995loi}.
Our calculation yields a Bayes factor of $2.2$ in favor of the string cosmology model over the power-law model. This suggests that the string cosmology model, which is based on string theory, provides a slightly better fit to the NANOGrav data than the SMBHB model. However, it is important to note that a Bayes factor of $2.2$ is considered to be only weak evidence in favor of the string cosmology model~\cite{Kass:1995loi}, and further data from future PTA observations will be necessary to confirm or refute this preference.

%%%%%%%%%%%%%%%%%%%%%%%%%%%%%%%%%%%%%%%%%%%%%%%%%%%%%%%%%%%%%%%%%%%%%%%%%%%%%%%%%%%%%%%%%%%%%%%%
\section{\label{conclusion}Conclusion}

In this work, we have utilized the NANOGrav 15-year data set to constrain the parameters of a string cosmology model for the SGWB. Our analysis focused on three key parameters: the frequency $f_s$ and the fractional energy density $\Omega_\mathrm{gw}^{s}$ of GWs generated at the end of the dilaton-driven stage, and the Hubble parameter $H_r$ at the end of the string phase.

By employing Bayesian parameter estimation techniques, we have obtained constraints on $f_s$ and $\Omega_\mathrm{gw}^{s}$ that are consistent with theoretical predictions from string cosmology. Specifically, we find $f_s = 1.2^{+0.6}_{-0.6}\times 10^{-8} \mathrm{Hz}$ and $\Omega_\mathrm{gw}^{s} = 2.9^{+5.4}_{-2.3}\times 10^{-8}$. These results demonstrate the ability of PTAs to probe the early Universe and constrain cosmological models.
However, our analysis reveals that the current NANOGrav data is not sensitive to the $H_r$ parameter, which remains unconstrained and follows the prior distribution. This finding highlights the need for future PTA observations with improved sensitivities to shed light on the reheating history of the Universe.

To assess the relative performance of our string cosmology model compared to alternative explanations for the NANOGrav signal, we have calculated the Bayes factor between the string cosmology model and a simple power-law model, which is commonly associated with the SGWB from SMBHBs. The Bayes factor of $2.2$ in favor of the string cosmology model suggests that string theory-based models may provide a better explanation for the NANOGrav signal than SMBHBs. However, this evidence is considered weak, and further data from future PTA observations will be necessary to confirm or refute this preference conclusively.

In conclusion, our results demonstrate the potential of PTAs to constrain cosmological models and study the early Universe. The constraints obtained on the string cosmology parameters $f_s$ and $\Omega_\mathrm{gw}^{s}$ represent a significant step forward in the search for observational signatures of string cosmology. As PTAs continue to improve their sensitivities and gather more data, we can expect even more stringent constraints on these parameters and a deeper understanding of the nature of the SGWB and the early Universe.

%%%%%%%%%%%%%%%%%%%%%%%%%%%%%%%%%%%%%%%%%%%%%%%%%%%%%%%%%%%%%%%%%%%%%%%%%%%%%%%%%%%%%%%%%%%%%%%%
\section*{Acknowledgments}
QT is supported by the National Natural Science Foundation of China (Grants No.~12405055 and No.~12347111), the China Postdoctoral Science Foundation (Grant No.~2023M741148), the Postdoctoral Fellowship Program of CPSF (Grant No. GZC20240458), and the innovative research group of Hunan Province under Grant No.~2024JJ1006.
YW is supported by the National Natural Science Foundation of China under Grant No.~12405057.
LL is supported by the National Natural Science Foundation of China Grant under Grant No.~12433001.  

\bibliographystyle{JHEP}
\bibliography{ref}
\end{document}